\documentclass[useAMS,usenatbib]{mn2e}
\usepackage{graphicx,amsmath,multirow,amssymb}
\usepackage{natbib}
\newcommand{\comment}[1]{}

\def\simgt{\lower.5ex\hbox{$\; \buildrel > \over \sim \;$}}
\def\simlt{\lower.5ex\hbox{$\; \buildrel < \over \sim \;$}}

\title{Dissecting the {\it Spitzer} color-magnitude diagrams of extreme LMC AGB stars}
\author[Dell'Agli et al.]{F. Dell'Agli$^{1,2}$, P. Ventura$^1$, D. A. Garc\'{\i}a Hern\'andez$^{3,4}$,
R. Schneider$^1$, 
\newauthor
M. Di Criscienzo$^{1}$, E. Brocato$^1$, F. D'Antona$^{1}$,  C. Rossi$^{2}$ \\
$^1$INAF -- Osservatorio Astronomico di Roma, Via Frascati 33, 00040, Monte Porzio Catone (RM), Italy \\
$^2$Dipartimento di Fisica, Universit\`a di Roma ``La Sapienza'', P.le Aldo Moro 5, 00143, Roma, Italy \\
$^{3}$Instituto de Astrof\'{\i}sica de Canarias, V\'{\i}a L\'actea s/n, E-38200 La Laguna, Tenerife, Spain \\
$^{4}$Dept. Astrof{\'{\i}}sica, Universidad de La Laguna (ULL), E-38206 La Laguna, Tenerife, Spain
}

\begin{document}

\date{Accepted, Received; in original form }

\pagerange{\pageref{firstpage}--\pageref{lastpage}} \pubyear{2012}

\maketitle

\label{firstpage}

\begin{abstract}
We trace the full evolution of low- and intermediate-mass stars  ($1 M_{\odot}
\leq M \leq 8M_{\odot}$) during the Asymptotic Giant Branch (AGB) phase in the
{\it Spitzer} two-color and color-magnitude diagrams. We follow the formation
and growth of dust particles in the circumstellar envelope with an isotropically
expanding wind, in which gas molecules impinge upon pre--existing seed nuclei,
favour their growth. These models are the first able to identify the main
regions in the {\it Spitzer} data occupied by AGB stars in the Large Magellanic
Cloud (LMC). The main diagonal sequence traced by LMC extreme stars in the
[3.6]-[4.5] vs. [5.8]-[8.0] and [3.6]-[8.0] vs. [8.0] planes  are nicely fit by
carbon stars models; it results to be an evolutionary sequence with the reddest
objects being at the final stages of their AGB evolution. The most extreme
stars, with $[3.6]-[4.5] > 1.5$ and $[3.6]-[8.0] > 3$, are $2.5-3 M_{\odot}$
stars surrounded by solid carbon grains. In higher mass ($>3 M_{\odot}$) models
dust formation is driven by the extent of Hot Bottom Burning (HBB) - most of the
dust formed is in the form of silicates and the maximum obscuration phase by
dust particles occurs when the HBB experienced is strongest, before the mass of
the envelope is considerably reduced. 
\end{abstract}


\section{Introduction}

A reliable estimate of the nature and the amount of dust produced by Asymptotic Giant
Branch (AGB) stars proves essential for a number of scientific issues. These stars are
believed to be the  dominant stellar sources of dust in the present-day Universe and
their contribution to dust enrichment can not be neglected even at redshift $z > 6$
\citep{valiante11}. In order to properly include their contribution in
chemical evolution models with dust, the mass and composition of dust grains released
by each star as a function of its mass and metallicity need to be known. In addition,
the corresponding size distribution  function allows to compute the extinction
properties associated with these grains, which is fundamental information required
for correctly interpreting the optical-near infrared  properties of high-z quasar and
gamma ray burst spectra  \citep{gallerani10}. 

Theoretical modelling of dust formation around AGBs has made considerable steps
forward in the last years, owing to the pioneering explorations by the Heidelberg
group \citep[see e.g.][and references therein]{fg06}, complemented by recent
studies \citep{paperI, paperII, paperIII, paperIV,nanni13a, nanni13b}. The reliability of
these models must be tested against the observations, given the many uncertainties
affecting the AGB phase modelling \citep[see e.g.][]{herwig05}, and the description of
the formation and growth of dust grains in the winds of AGBs \citep{fg06}.

On the observational side, the study of evolved stars in the Galaxy is hampered by the
obscuration determined by the interstellar medium, and by the unknown distances of the
objects observed. This pushed the interest towards other nearby galaxies. The Large
Magellanic Cloud (LMC) is an optimum target for this scope. This stems from the low
average reddening ($E(B-V) \sim 0.075$) and its proximity, that allows the
determination of important stellar properties, such as the absolute magnitudes. The LMC
has been surveyed in the optical by the Magellanic Clouds Photometric Survey
\citep[MCPS,][]{zaritsky97}, in the near-IR by the Deep Near Infrared Survey of
the Southern Sky \citep[DENIS,][]{epchtein94} and the Two Micron All Sky Survey
\citep[2MASS,][]{skrutskie06}. Of particular interest for the study of evolved,
dust--surrounded stars, is the SAGE (Surveying the Agents of Galaxy Evolution)
survey obtained by the {\it Spitzer Space Telescope} with the Infrared Array Camera
(IRAC; 3.6, 4.5, 5.8 and 8.0 $\mu$m) and the Multiband Imaging Photometer for {\it
Spitzer} (MIPS; 24, 70 and 160 $\mu$m) \citep[see e.g.][]{meixner06}. The analysis of
evolved stars based on 2MASS and IRAC data lead to a classification of LMC stars into
three main categories. \citet{cioni06}, based on their analysis of the 2MASS
color-magnitude diagram (CMDs), divided the region above the tip of the Red Giant
Branch (RGB) into O-rich  and C-rich zones; \citet{lmcpaperII} selected a group of
stars, called ``extreme'', showing the clear signature of the presence of a dusty
circumstellar envelope. These stars were shown to contribute about 75\% of the
overall dust from AGBs (Riebel et al. 2012). This stimulated an interesting series of
papers \citep{lmcpaperI, lmcpaperIV, lmcpaperV, lmcpaperVI} aimed at
refining such a classification, to interpret the observed CMDs by using grids of
synthetic spectra for various chemical and physical inputs, the most relevant being the
surface chemistry of the star, the size of the grains formed, the borders of the dusty
shell, the surface gravity and the effective temperature of the central object.

In this Letter we use models of dust formation around AGBs,  based on full
evolutionary computations, to interpret the distribution of LMC extreme stars in the
color--color and color--magnitude diagrams (CCDs and CMDs, respectively). Our
goal is to characterize the extreme stars in terms of the evolutionary phase, and of
the amount and type of dust present in their surroundings. This investigation
will be an important benchmark for the studies focused on dust formation around AGB
stars.

\section{Stellar evolution and dust formation modelling}

\begin{figure*}
\begin{minipage}{0.33\textwidth}
\resizebox{1.\hsize}{!}{\includegraphics{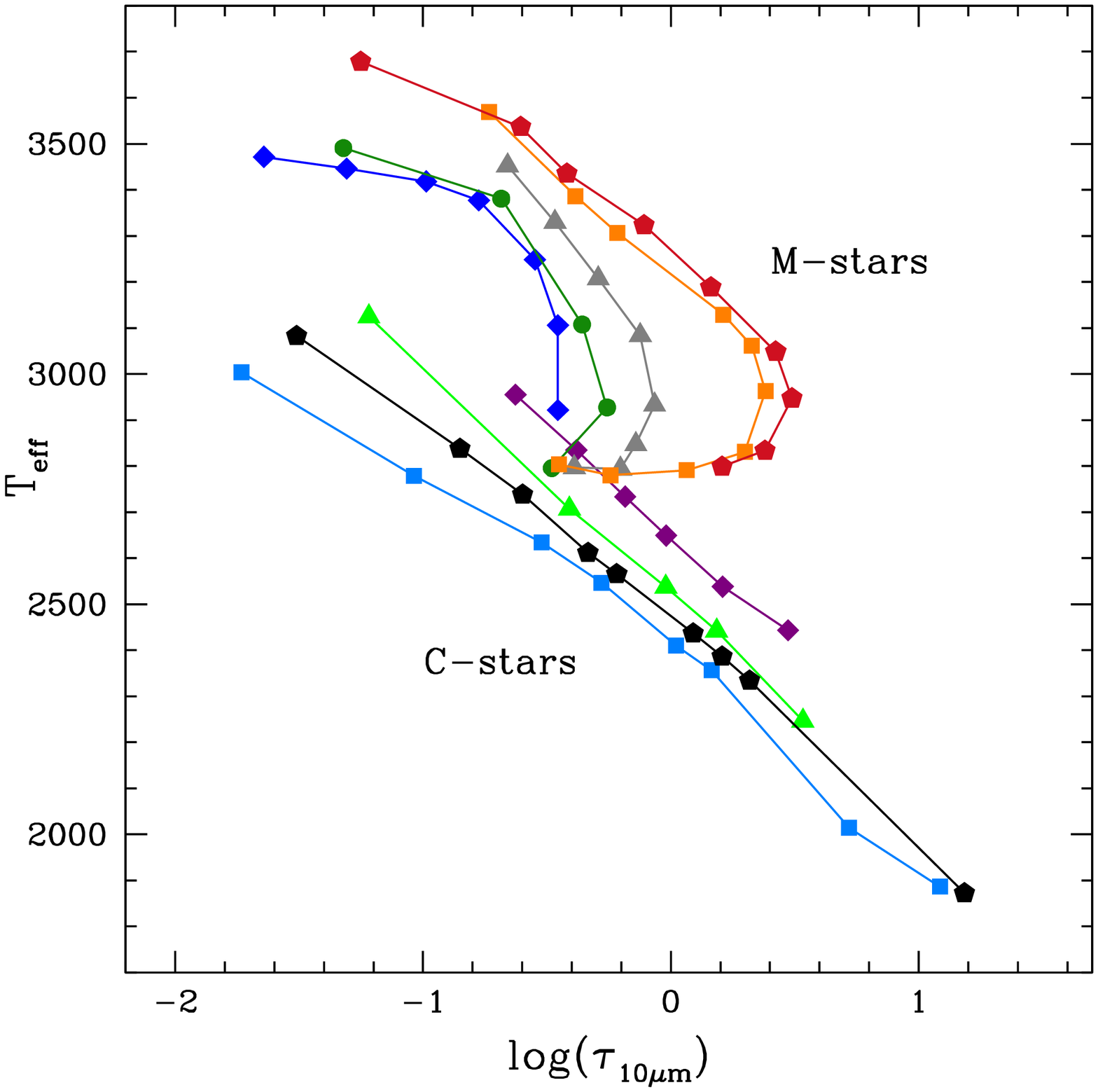}}
\end{minipage}
\begin{minipage}{0.33\textwidth}
\resizebox{1.\hsize}{!}{\includegraphics{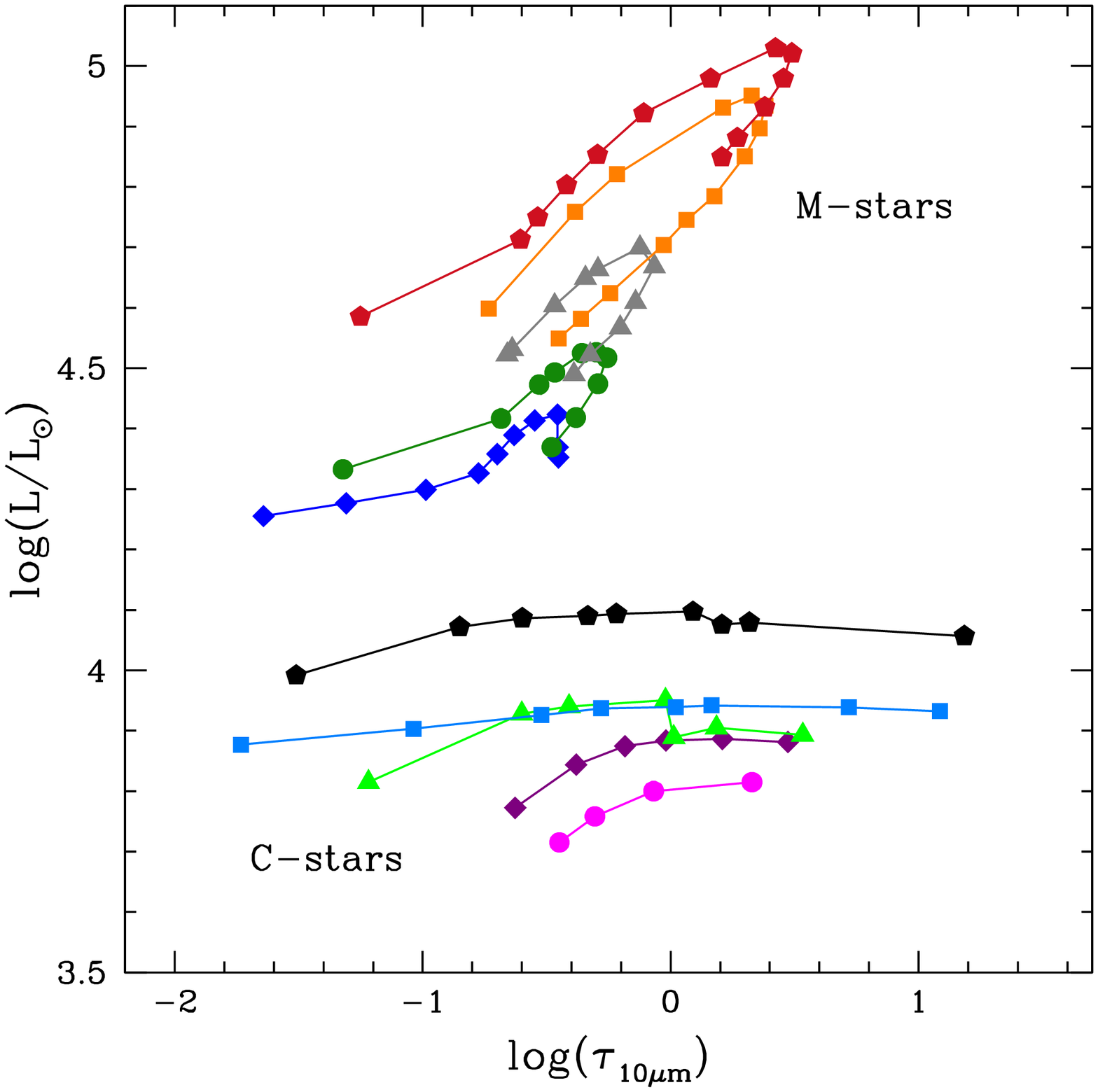}}
\end{minipage}
\begin{minipage}{0.33\textwidth}
\resizebox{1.\hsize}{!}{\includegraphics{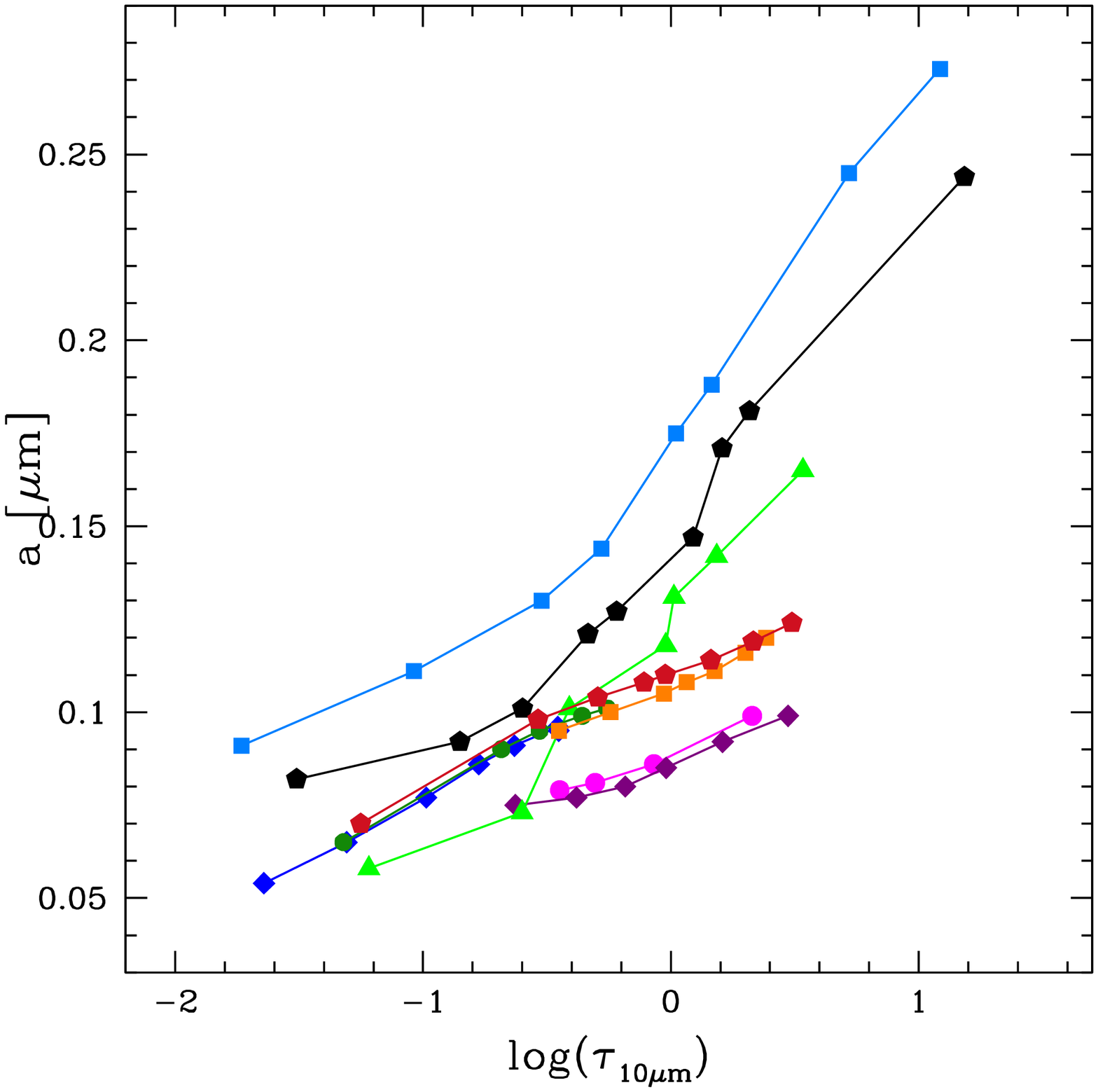}}
\end{minipage}
\vskip+10pt
\caption{The evolution of AGB models of various mass shown as a function of $\tau_{10\mu m}$,
indicating the degree of obscuration. The panels show the variation of the effective
temperature (left panel), luminosity (middle) and size of the grains formed (right); 
we refer to olivine for M--stars, and solid carbon for C--stars). The meaning of the 
symbols is as follows. Magenta circles: $1.25M_{\odot}$; violet diamonds: $1.5M_{\odot}$; 
light green triangles: $2M_{\odot}$; light blue squares: $2.5M_{\odot}$; black pentagons: 
$3M_{\odot}$; blue diamonds: $3.5M_{\odot}$; green circles: $4M_{\odot}$; grey triangles: 
$5M_{\odot}$; orange squares: $6.5M_{\odot}$; red pentagons: $7.5M_{\odot}$.
}
\label{fevol}
\end{figure*}

\begin{figure*}
\begin{minipage}{0.47\textwidth}
\resizebox{1.\hsize}{!}{\includegraphics{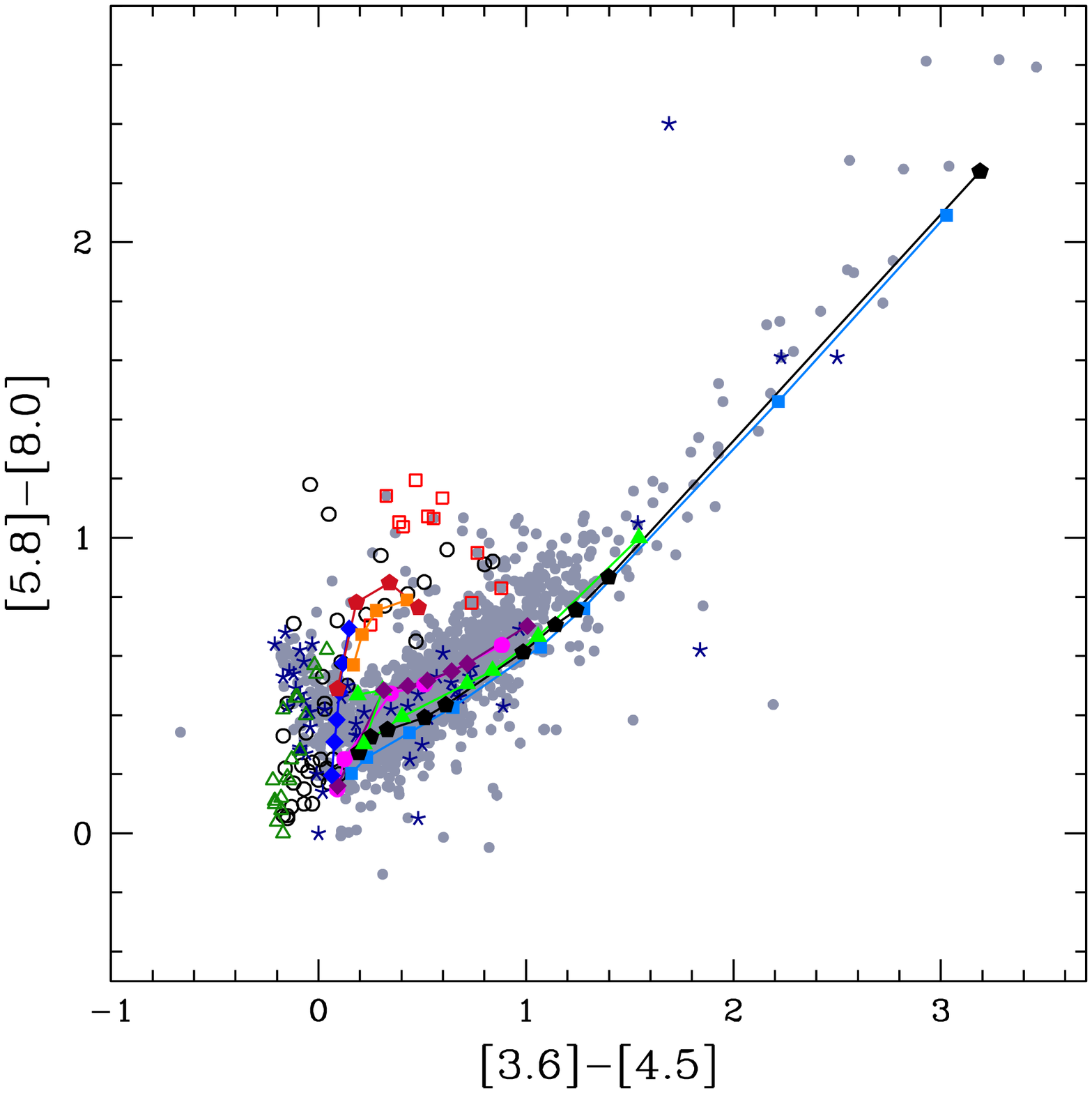}}
\end{minipage}
\begin{minipage}{0.47\textwidth}
\resizebox{1.\hsize}{!}{\includegraphics{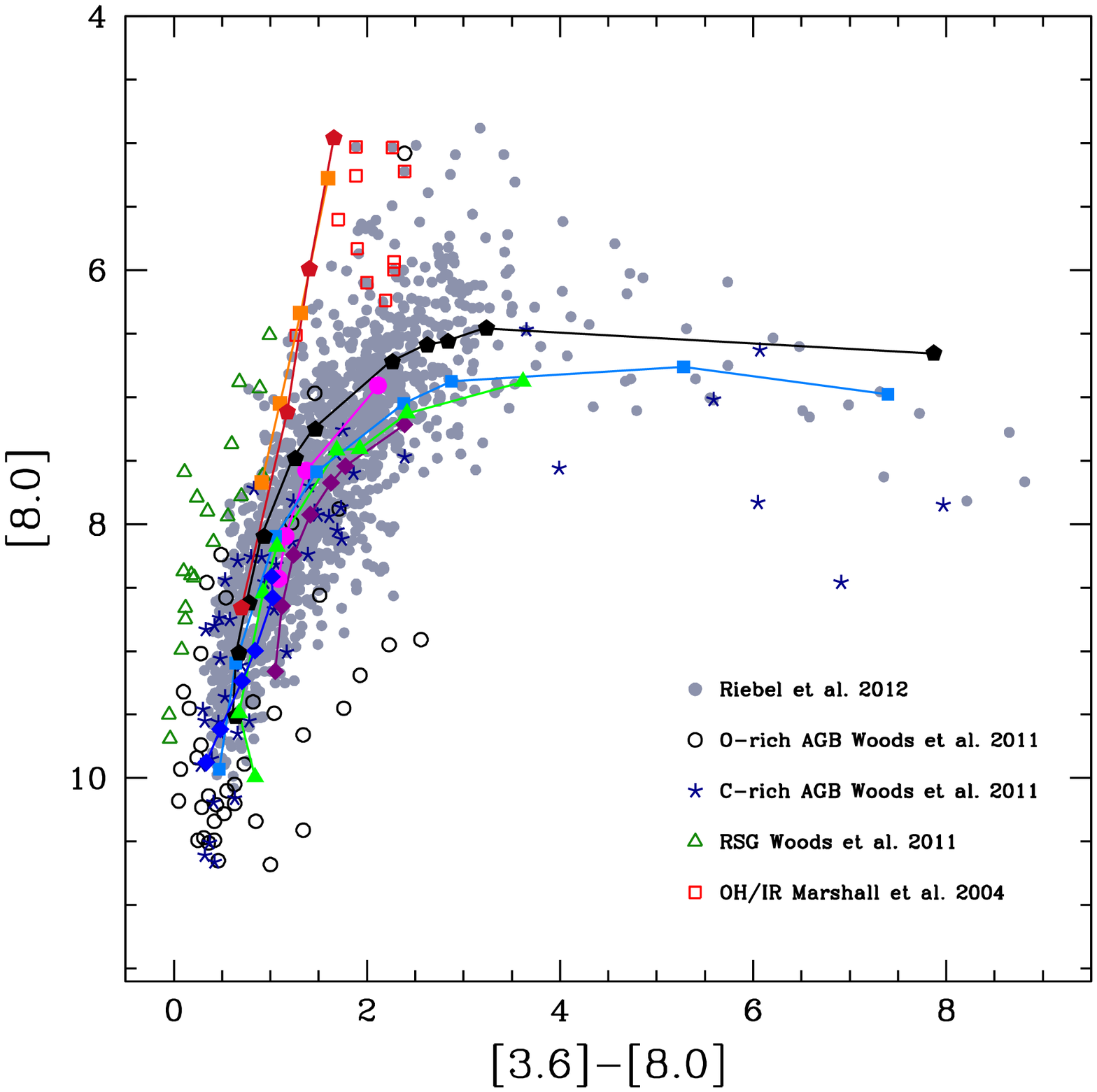}}
\end{minipage}
\vskip+10pt
\caption{Data of LMC extreme stars taken from \citet{riebel12} are shown as
solid, grey circles in the $[3.6]-[4.5]$ vs $[5.8]-[8.0]$ (left) and $[3.6]-[8.0]$ vs  $[8.0]$ (right) diagrams. 
The position of the models during the AGB
evolution are also shown (colors are the same as in Fig. \ref{fevol}).
Spectroscopically confirmed C- and O-rich AGBs and RSG stars (from Woods et al.
2011) as well as extreme OH/IR stars (from Marshall et al. 2004) are also
marked with different symbols/colors in these diagrams.}
\label{fcol}
\end{figure*}

The evolutionary sequences have been calculated by means of the ATON code for
stellar evolution. The interested reader may find the details of the
numerical structure of the code together with the more recent updates in
\citet{ventura09}. The range of masses investigated is $ 1M_{\odot} \leq M \leq
8M_{\odot}$. More massive stars undergo core collapse by  e--capture, and do not
experience the AGB phase. Models with  $6.5 M_{\odot} < M < 8 M_{\odot}$ achieve
carbon ignition in conditions of partial degeneracy, and develop a core made up
of oxygen and neon; their phase of thermal pulses is commonly known as super-AGB
\citep[hereafter SAGB,][]{siess06}. We assumed a chemical composition typical
for LMC stars: $Z=0.008$, $Y=0.26$. 

We recall here the physical inputs most relevant to the results obtained. The 
convective instability is treated according to the Full Spectrum of Turbulence (FST) 
description, developped by \citet{cm91}; in models whose initial mass exceeds $\sim 3
M_{\odot}$, use of  the FST model leads to strong Hot Bottom Burning
\citep[HBB,][]{renzini81} conditions, with the activation of an advanced p--capture 
nucleosynthesis at the bottom of the convective envelope \citep{vd05}. Mass loss for
M--stars is modelled following \citet{blocker95}; for carbon stars, we use the results
by \citet{wachter08}. In regions unstable to convective motions, nuclear burning and
mixing of chemicals are coupled via a diffusive approach, as described in
\citet{cloutmann}; during the AGB phase, convective velocities are allowed to decay
exponentially into radiatively stable regions, based on the calibration aimed at
reproducing the observed luminosity function of carbon stars, discussed in
\citet{paperIV}. The wind surrounding AGBs is assumed to expand isotropically, and to
be accelerated by radiations pressure on dust particles. The growth of dust grains is
described by condensation of gas molecules in the wind impinging on the already
formed grains. The relevant equations can be found, e.g., in \citet{fg06}, and  were
already used in other investigations by our group \citep{paperI, paperII, paperIII,
paperIV}, and also by \citet{nanni13a, nanni13b}.

The spectral energy distribution (SED) of the star during the various
evolutionary phases was determined from the knowledge of the properties of the central
object and of the structure of the dust shell by means of the DUSTY code \citep{dusty}.
 For M-stars we considered the presence of amorphous silicates (mainly 
olivine) and alumina dust grains, whereas solid amorphous carbon and 
SiC particles were used for C-stars. Iron grains are also accounted 
for in both cases. In addition, for a given AGB evolutionary time, the 
location (and temperature) of the inner dust region boundary, the 
density stratification, the relative percentages of the different dust 
species (i.e., SiC vs. AmC, Silicates vs. alumina), the optical depth 
and the dust grain size as calculated self-consistently by our wind 
AGB models (Ventura et al. 2014 and references therein) are then used 
as inputs in DUSTY (see Dell'Agli et al., in preparation). Finally, to 
test the reliability of the results obtained, we calculated a few SEDs 
with the 2-DUST code (Ueta \& Meixner 2003), run in the spherically 
symmetric mode. The location of the various dusty layers and the 
relative percentages of the dust species present are assumed 
consistently with the results from the wind modelling. These tests 
showed no differences ($< $10\% in flux) among the results obtained with 
the two codes.

\section{Dust formation during the AGB phase}

Fig. \ref{fevol} shows the evolution of models of different mass during the AGB phase.
Because this work is focused on the most obscured objects, we restrict our analysis to
stars that have already reached the C-star stage, or M-stars where HBB is active;
these are the only conditions allowing dust production in meaningful quantities. We
show the variation of the effective temperature, luminosity and size of the grains
formed as a function of the optical depth at 10 $\mu$m ($\tau_{10 \mu{\rm m}}$),
which quantifies the degree of obscuration of the central star. It is not possible to
define a unique grain size, because particles of various species form, with different
dimensions: here we refer to the grains that dominate the dust mass formed, providing
most of the acceleration of the wind by radiation pressure, i.e. solid carbon for
C-stars and olivine for M-stars \citep{fg06}.

The evolution of C-stars is driven by the carbon enrichment at the surface, favoured
by the occurrence of Third Dredge Up (TDU). As the surface of the star becomes more
enriched in carbon, the star assumes a more extended configuration \citep{marigo02,
vm09, vm10}, evolves towards lower effective temperatures (T$_{eff}$) and loses
mass at larger rates \citep{wachter08}: all these factors favour dust production,
mainly under the form of solid carbon, with traces of SiC and iron \citep{fg06}.
We see in the left panel of Fig. \ref{fevol} that the stars become more and more
obscured as the external layers cool; C-stars with the same T$_{eff}$ have the
same degree of obscuration, independent of their previous history. The most obscured
stars, with $\tau_{10 \mu{\rm m}} \sim 10$, are those in the final stages of the AGB
evolution, when the size of the carbon grains is $a_C \sim 0.20-0.25 \mu$m
\citep{paperIV, nanni13a}. The
middle panel of Fig. \ref{fevol} shows that the luminosity has only a minor effect on
the optical depth variation of C-stars; higher mass stars evolve at larger
luminosities, owing to their larger core masses. We note that these optical depth 
(and luminosity) values are consistent with
those derived for the most extreme C-stars in the LMC (Gruendl et al. 2008).

Stars with initial mass above $3 M_{\odot}$ experience HBB, that prevents the
possibility of reaching the C-star stage; their evolution is driven by the temperature
at the bottom of the envelope, which determines the luminosity, the mass-loss
rate and consequently the quantity of dust formed. As shown in the middle panel of
Fig. \ref{fevol}, the luminosity ($L$) undergoes significant variations during the AGB
evolution:  after the initial increase, due to the growth of the core mass, $L$
decreases, owing to the gradual loss of the external mantle. The phase of maximum dust
production (with the highest $\tau_{10 \mu{\rm m}}$) is in conjunction with the
maximum $L$, when the mass--loss rate alsoattains its largest values. Note that L 
in extreme OH/IRs ($10^4-10^5 L_{\odot}$; Marshall et al. 2004) agrees very well with our 
predictions for the more massive AGBs. Stars in the
SAGB regime, owing to their large core mass, reach the highest values of $\tau_{10
\mu{\rm m}}$, that approaches $\sim 3$ in the $7.5 M_{\odot}$ model. The right panel
of Fig. \ref{fevol} shows the tight relationship between the size of the olivine
grains ($a_{\rm ol}$) formed and $\tau_{10 \mu{\rm m}}$; $a_{\rm ol}$ changes from
$\sim 0.1 \mu$m to $\sim 0.13 \mu$m as $\tau_{10 \mu{\rm m}}$ grows from 1 to $\sim
3$. In the high-mass domain the T$_{eff}$ has a minor effect on $\tau_{10
\mu{\rm m}}$, in agreement with the fact that dust production is
mostly determined by the extent of the HBB experienced. The narrow $T_{eff}$ range
($2700 < T_{eff} < 3500$ K) predicted in the more massive AGB stars agree quite well
with the spectroscopic temperatures derived for such stars in the LMC \citep{garcia09}
and our own Galaxy \citep{garcia06, garcia07}\footnote{Note, however, that C-rich AGB
stars usually display much lower effective temperatures \citep[e.g.][]{abia01} as our 
models predict.}.

\section{The interpretation of the extreme stars in the LMC}

The {\it Spitzer} data of extreme stars in the LMC from Riebel et al.
(2012) are shown in the two panels of Fig. \ref{fcol}. We also display the
smaller group of spectroscopically confirmed C- and O-rich AGBs (and luminous
red supergiants, RSGs) from Woods et al. (2011) as well as the small sample of
extreme OH/IR stars from Marshall et al. (2004)\footnote{ These are expected
to be the most massive LMC AGB stars \citep{garcia09}.}. The position of our
models during different stages of the AGB phase are also indicated.

In the $[5.8]-[8.0]$ vs. $[3.6]-[4.5]$ plane the theoretical models of C-stars define
a diagonal sequence, that nicely fits the observed colors. We interpret the reddest
objects, with  $[3.6]-[4.5] > 1.5$, as the latest evolutionary stages of stars of
initial mass $\sim 2.5-3 M_{\odot}$, just below the threshold to ignite HBB. These are
the only progenitor masses in which the surface C reaches abundances $X(C) \sim
0.01$, with a great production of carbon dust, provoking a strong obscuration of
the central star. We expect that only a few stars populate this region, as confirmed
by the observations.

The observed spread in $[5.8]-[8.0]$ for a given value of $[3.6]-[4.5]$ is due to the 
different relative distribution of SiC and carbon dust. Models with a high SiC/C ratio
exhibit a more prominent feature at $11.3 \mu$m, and will be less obscured,
particularly in the $[3.6]-[4.5]$ color. This is the case of stars with mass $M \leq
1.5 M_{\odot}$, that will populate the upper region in the color-color plane at
$[3.6]-[4.5] < 1.5$. Strong production of carbon dust limits the formation of SiC,
owing to the strong acceleration of the wind. Therefore, the stars whose SED exhibits
the most pronounced feature of SiC are those of initial mass below $\sim 2 M_{\odot}$,
in which the surface C abundance keeps below $X(C) \sim 0.01$. More massive
models ($M\sim 2.5-3 M_{\odot}$) achieve extremely high surface C abundances. In the
latest AGB phases, these models evolve at $[3.6]-[4.5] > 1.5$ and the SED is
characterized by the appearance of a weak absorption SiC feature at $11.3 \mu m$.
Thus, our finding confirms the interpretation/observations of the LMC ``extremely red
objects'' by Gruendl et al. (2008).

C-stars models also fit the observed {\it Spitzer} CMD. The observed spread in the $8 \mu$m
flux is a luminosity effect; higher mass stars evolve on more massive cores, thus at larger
luminosities (see middle panel of fig. \ref{fevol}), and smaller $[8.0]$ magnitudes. In
both panels of Fig.2 we see that the position of the spectroscopically confirmed C-rich AGBs
by Woods et al. (2011) is nicely reproduced by our models. The only exception is the group of
objects at $[3.6]-[4.5] < 0$, representing the C-stars with little (if any) dust in their
envelopes. Their {\it Spitzer} colors and magnitudes are reproduced by the dust-free
photosphere models by Aringer et al. (2009).

AGB stars experiencing HBB trace a different path. As shown in fig. \ref{fevol},
these stars evolve at smaller optical depths compared to C-stars, because the
extinction coefficient of carbon dust exceeds that of silicates. In the CCD,
the  theoretical tracks are confined in the region $[3.6]-[4.5] < 0.5$. The slope of
the line traced by the theoretical tracks is different from that traced by
C-stars; {this is because the strong silicates feature} determines a great increase in
the $8 \mu$m  flux, with a considerable increase in the $[5.8]-[8.0]$ color. 

Massive AGBs with $M > 6M_{\odot}$ evolve at $\tau_{10 \mu m} > 2$ (see fig.
\ref{fevol}), and populate the region at $[5.8]-[8.0] > 0.5$. Unlike their lower mass
counterparts, the  reddest colors are not reached in the final stages of the AGB
evolution, but during the phase of maximum luminosity, when the HBB experienced is
strongest. In the CMD, most of the M-stars populate the regions at $[3.6]-[8.0]
< 1$, and mix with the C-stars. The two sequences bifurcate at $[8.0] \sim 8$, because
the sequence of massive AGBs, owing to the effects of the silicates feature, trace a
steep diagonal line, up to $[8.0] \sim 5$. Interestingly, the extreme LMC OH/IR
stars (very massive AGBs) are located at the highest [8.0] magnitudes in good
agreement with our massive models; indeed the four OH/IR stars closer to the model
predictions are confirmed very massive Rb-rich LMC AGBs \citep{garcia09}. Models of
mass above $6 M_{\odot}$ experience strong HBB, and evolve at luminosities above the
classic limit for AGBs, $L \sim 5\times 10^4 L_{\odot}$  (i.e. $M_{bol}=-7.1$).

 In short, our interpretation of the extreme stars in the LMC is in good
agreement with the observations of spectroscopically confirmed C- and O-rich LMC
AGBs by \citet{woods11}. The only difference is in the identification of
high luminosity stars with $[8.0] \sim 5 - 6$, that we interpret as massive AGBs
experiencing strong HBB, whereas Woods et al. (2011) identify as RSGs. Our work
suggests that RSGs may be distinguished from massive AGBs/SAGBs in the CMD {\it
Spitzer} diagram but not in the CCD one (see Fig. 2); indeed a small fraction of
RSGs may be truly massive  AGBs/SAGBs. The O-rich sample by Woods et al. (2011)
is nicely reproduced by our massive AGB models in the two {\it Spitzer} diagrams
(Fig. 2), with the exception of the the bluest objects, at $[3.6]-[4.5] < 0$
that, like their C-rich counterparts, have almost dust-free envelopes.

\section{Conclusions}

 For the first time, we are able to interpret the {\it Spitzer} observations of
extreme stars in the LMC on the basis of evolutionary models of dusty AGB stars. 

We find that the main diagonal sequence traced by the observations in the $[5.8]-[8.0]$
vs. $[3.6]-[4.5]$ diagram is an evolutionary sequence of C-stars that become
progressively more obscured as their surface layers dredge-up carbon from the stellar
interior. We identify the reddest objects, with $[3.6]-[4.5] > 1.5$, as the descendants of
$\sim 2.5-3 M_{\odot}$ stars in the latest evolutionary phases, when they develop an optically
thick circumstellar envelope. The distribution of stars in the {\it Spitzer} CCD is generally
determined by the optical depth, but it is also dependent on the relative percentages of
carbon and SiC dust formed. Stars with a high SiC/carbon ratio populate the upper region of
the diagram. This motivates the spread in  $[5.8]-[8.0]$ observed for $[3.6]-[4.5] < 1.5$; the
scatter disappears for more obscured objects, because the dust formed is entirely dominated by
solid carbon. 

Massive AGBs with $M > 3 M_{\odot}$ experience HBB, which prevents them from becoming
C-stars; in their circumstellar envelopes the formation of silicates occurs. The strong
feature of silicates at $9.7 \mu$m favours a considerable increase in the $8.0 \mu$m flux,
thus the slope traced by the evolutionary sequences in the {\it Spitzer} CCD is
different from C-stars. Owing to the small extinction coefficients of silicates, M-stars do
not reach extremely red colors; most of dusty massive AGBs are confined in the region
$[3.6]-[4.5] < 0.5$, also populated by C-stars. The most massive AGBs experience a phase
of strong HBB, when the production of silicates is strongly enhanced: both in the  {\it
Spitzer} CCD and CMD these stars will evolve above the main sequence, where most of the
stars are detected. 

 Follow-up (optical and near-IR) spectroscopic observations of the extreme
stars in the LMC will be extremely useful to confirm our analysis.

\section*{Acknowledgments}
The authors are indebted to the anonymous referee for the careful reading 
of the manuscript and for the detailed and relevant comments, that 
helped to increase the quality of this work.
D.A.G.H. acknowledges support provided by the Spanish Ministry of Economy and
Competitiveness under grants AYA$-$2011$-$27754 and AYA$-$2011$-$29060.
P.V. was supported by PRIN MIUR 2011 "The Chemical and Dynamical Evolution of the Milk Way and Local Group Galaxies" (PI: F. Matteucci), prot. 2010LY5N2T.
R.S. acknowledges funding from the European Research Council under the European Union's
Seventh Framework Programme (FP/2007-2013) / ERC Grant Agreement n. 306476.

\end{document}